\documentstyle[amssymb,floats,aps,prb,epsf]{revtex}
\epsfclipon
\begin{document}
\twocolumn[\hsize\textwidth\columnwidth\hsize\csname
@twocolumnfalse\endcsname

\draft
\title{A gauge invariant dressed holon and spinon description of
the normal-state of underdoped cuprates}
\author{Shiping Feng, Jihong Qin, and Tianxing Ma}
\address{Department of Physics, Beijing Normal University, Beijing
100875, China}


\maketitle

\begin{abstract}
A partial charge-spin separation fermion-spin theory is developed
to study the normal-state properties of the underdoped cuprates.
In this approach, the physical electron is decoupled as a gauge
invariant dressed holon and spinon, with the dressed holon
behaving like a spinful fermion, representing the charge degree of
freedom together with the phase part of the spin degree of
freedom, while the dressed spinon is a hard-core boson,
representing the amplitude part of the spin degree of freedom. The
electron local constraint for the single occupancy is satisfied.
Within this approach, the charge and spin dynamics of the
underdoped cuprates are studied based on the $t$-$t'$-$J$ model.
It is shown that the charge dynamics is mainly governed by the
scattering from the dressed holons due to the dressed spinon
fluctuation, while the scattering from the dressed spinons due to
the dressed holon fluctuation dominates the spin dynamics.
\end{abstract}
]
\bigskip
\narrowtext

\section{Introduction}

After over fifteen years of intense experimental studies of doped
cuprate superconductors, a significant body of reliable and
reproducible data has been accumulated by using many probes
\cite{kbse1,bcmps}, which shows that the most remarkable
expression of the nonconventional physics is found in the
normal-state \cite{kbse1,bcmps}. The normal-state properties
exhibit a number of anomalous properties in the sense that they do
not fit in the conventional Fermi-liquid (FL) theory
\cite{bcmps,pwa1,pwa2}, and are closely related to the fact that
these materials are doped Mott insulators
\cite{kbse1,bcmps,pwa1,pwa2}. The ground state in the undoped case
is an antiferromagnetic (AF) long-range-order (AFLRO) N\'eel
state, but changing the carrier concentration by ionic
substitution or increasing the oxygen content turns these
compounds into strongly correlated metals leaving the AF
short-range-order correlation still intact \cite{kbse1,bcmps}. It
is then not surprising that the nonconventional behaviors are most
striking in the underdoped regime, where the concentration of
doped holes is small. The anomalous properties observed in a
variety of experiments, such as the nuclear magnetic resonance
(NMR), nuclear quadrupole resonance (NQR), muon spin rotation
($\mu SR$) techniques, inelastic neutron scattering studies
\cite{rossat,kbse2,kbse3,dai1,kbse4,kbse5}, optical and transport
measurements \cite{cooper1,uchida1,uchida2,ando1}, and
angle-resolved photoemission spectroscopy (ARPES) investigation
\cite{shen1}, exclude conventional theories. The single common
feature in doped cuprates is the two-dimensional (2D) CuO$_{2}$
plane \cite{kbse1,bcmps}, and it seems evident that the
nonconventional behaviors are dominated by this plane. Very soon
after the discovery of high-temperature superconductivity (HTSC)
in doped cuprates, Anderson proposed a scenario of HTSC based on
the charge-spin separation (CSS) in 2D \cite{pwa1}, where the
internal degrees of freedom of the electron are decoupled as the
charge and spin degrees of freedom, while the elementary
excitations are not quasi-particles but collective modes for the
charge and spin degrees of freedom, i.e., the holon and spinon,
then these holon and spinon might be responsible for the
nonconventional behaviors. Many unusual properties of the
underdoped cuprates are extensively studied following this line
within the 2D $t$-$J$ type model \cite{pwa2,dagotto1}.

The decoupling of the charge and spin degrees of freedom of
electron is undoubtedly correct in one-dimensional (1D)
interacting electron systems \cite{haldane}, where the charge and
spin degrees of freedom are represented by boson operators that
describe the excitations of charge-density wave and spin-density
wave, respectively. In particular, the typical behavior of the
non-FL, i.e., the absence of the quasiparticle propagation and
CSS, has been demonstrated theoretically within the 1D $t$-$J$
model \cite{ogata}. Moreover, the holon and spinon as the real
elementary excitations in 1D cuprates have been observed directly
by the ARPES experiment \cite{shen2}. Therefore both theoretical
and experimental studies indicate that the existence of the real
holon and spinon is common in 1D \cite{maekawa}. However, the case
in 2D is very complex since there are many competing degrees of
freedom \cite{kbse1,bcmps}. As a consequence, both experimental
investigation and theoretical understanding are extremely
difficult \cite{kbse1,bcmps,pwa1,pwa2}. Among the unusual
properties of the underdoped cuprates, a hallmark is the charge
transport \cite{cooper1,uchida1,uchida2,ando1}, where the
conductivity shows a non-Drude behavior at low energies, and is
carried by $x$ holes, where $x$ is the hole doping concentration,
while the resistivity exhibits a linear temperature behavior over
a wide range of temperatures. This is a strong experimental
evidence supporting the notion of CSS, since not even conventional
electron-electron scattering would show the striking linear rise
of scattering rate above the Debye frequency, and if there is no
CSS, the phonons should affect these properties \cite{pwa4}.
Moreover, it has been argued that the most plausible source of the
absence of phonon scattering and of pair-breaking effects seems to
be CSS \cite{pwa4}, and further, a compelling evidence for CSS in
doped cuprates has been found from the experimental test of the
Wiedemann-Franz law, where a clear departure from the universal
Wiedemann-Franz law for the typical FL behavior is observed
\cite{hill}. On the numerical study front, the crossover from the
FL to non-FL behavior with decreasing the hole doping
concentration near the Mott transition has been found within the
2D $t$-$J$ model \cite{rice3}. Furthermore, it has been shown
within the 2D $t$-$t'$-$J$ model by using the exact
diagonalization method that there is a tendency of holes to
generate nontrivial spin environments, this effect leads to a
decoupling of the spin from charge \cite{dagotto2}. In this case,
a formal theory with the gauge invariant holon and spinon, i.e.,
the issue of whether the holon and spinon are real, is centrally
important \cite{laughlin}. In this paper, we propose a partial CSS
fermion-spin theory, and show that if the local single occupancy
constraint is treated properly, then the physical electron can be
decoupled completely by introducing the dressed holon and spinon.
These dressed holon and spinon are gauge invariant, i.e., they are
real in 2D. As an application of this theory, we discuss the
charge and spin dynamics of the underdoped cuprates within the
$t$-$t'$-$J$ model, and the results are qualitatively similar to
that seen in experiments.

The paper is organized as follows. The framework of the partial
CSS fermion-spin theory is presented in Sec. II. Within this
theory, the single-particle dressed holon and spinon Green's
functions of the $t$-$t'$-$J$ model are calculated in Sec. III by
considering the dressed holon-spinon interaction, where the dressed
holon and spinon self-energies are obtained by using the equation
of motion method. In Sec. IV, we discuss the charge transport of
the underdoped cuprates. The incommensurate (IC) spin response of
the underdoped cuprates is studied in Sec. V.  Sec. VI is devoted
to a summary and discussions.

\section{Gauge invariant dressed holon and spinon}

We start from the $t$-$t'$-$J$ model defined on a square lattice
as \cite{pwa1,rice1},
\begin{eqnarray}
H&=&-t\sum_{i\hat{\eta}\sigma}C^{\dagger}_{i\sigma}
C_{i+\hat{\eta}\sigma}+t'\sum_{i\hat{\tau}\sigma}
C^{\dagger}_{i\sigma}C_{i+\hat{\tau}\sigma} \nonumber \\
&+&\mu \sum_{i\sigma}C^{\dagger}_{i\sigma}C_{i\sigma}+
J\sum_{i\hat{\eta}}{\bf S}_{i}\cdot {\bf S}_{i+\hat{\eta}},
\end{eqnarray}
where $\hat{\eta}=\pm\hat{x},\pm\hat{y}$,
$\hat{\tau}=\pm\hat{x}\pm\hat{y}$, $C^{\dagger}_{i\sigma}$
($C_{i\sigma}$) is the electron creation (annihilation) operator,
${\bf S}_{i}=C^{\dagger}_{i}{\vec\sigma}C_{i}/2$ is spin operator
with ${\vec\sigma}=(\sigma_{x},\sigma_{y},\sigma_{z})$ as Pauli
matrices, and $\mu$ is the chemical potential. The $t$-$t'$-$J$
model (1) is subject to an important local constraint
$\sum_{\sigma}C^{\dagger}_{i\sigma}C_{i\sigma}\leq 1$ that a given
site can not be occupied by more than one electron. In the
$t$-$t'$-$J$ model, the strong electron correlation manifests
itself by this constraint, and therefore the crucial requirement
is to impose this constraint \cite{pwa1,rice1,rice2,rice6}. It has
been shown that this constraint can be treated properly within the
fermion-spin theory \cite{feng1}. In this section, we show that
the dressed holon and spinon in the fermion-spin theory are gauge
invariant. Following the discussions in [29], we decouple the
constrained electron operator as,
\begin{eqnarray}
C_{i\sigma}=h^{\dagger}_{i}a_{i\sigma},
\end{eqnarray}
with the constraint
$\sum_{\sigma}a^{\dagger}_{i\sigma}a_{i\sigma}=1$, where the
spinless fermion operator $h_{i}$ keeps track of the charge degree
of freedom, while the boson operator $a_{i\sigma}$ keeps track of
the spin degree of freedom, then the Hamiltonian (1) can be
rewritten as,
\begin{eqnarray}
H&=&-t\sum_{i\hat{\eta}\sigma}h_{i}a^{\dagger}_{i\sigma}
h^{\dagger}_{i+\hat{\eta}}a_{i+\hat{\eta}\sigma}+
t'\sum_{i\hat{\tau}\sigma}h_{i}a^{\dagger}_{i\sigma}
h^{\dagger}_{i+\hat{\tau}}a_{i+\hat{\tau}\sigma} \nonumber \\
&-&\mu\sum_{i}h^{\dagger}_{i}h_{i}+J\sum_{i\hat{\eta}}
(h_{i}h^{\dagger}_{i}){\bf S}_{i}\cdot{\bf S}_{i+\hat{\eta}}
(h_{i+\hat{\eta}}h^{\dagger}_{i+\hat{\eta}}),
\end{eqnarray}
where the pseudospin operator
${\bf S}_{i}=a^{\dagger}_{i}{\vec\sigma}a_{i}/2$. In this case, the
electron constraint $\sum_{\sigma}C^{\dagger}_{i\sigma}C_{i\sigma}
=1-h^{\dagger}_{i}h_{i}\leq 1$ is exactly satisfied, with
$n^{(h)}_{i}=h^{\dagger}_{i}h_{i}$ is the holon number at site $i$,
equal to 1 or 0. This decoupling scheme is called as the CP$^{1}$
representation \cite{ioffe}. The advantage of this formalism is
that the charge and spin degrees of freedom of the electron may be
separated at the mean field (MF) level, where the elementary charge
and spin excitations are called the holon and spinon, respectively.
We call such holon and spinon as the {\it bare holon and spinon},
respectively, since an extra $U(1)$ gauge degree of freedom related
with the constraint
$\sum_{\sigma}a^{\dagger}_{i\sigma}a_{i\sigma}=1$ appears, i.e.,
the CP$^{1}$ representation is invariant under a local $U(1)$ gauge
transformation,
\begin{eqnarray}
h_{i}\rightarrow h_{i}e^{i\theta_{i}}, ~~~
a_{i\sigma}\rightarrow a_{i\sigma}e^{i\theta_{i}},
\end{eqnarray}
and then all physical quantities should be invariant with respect
to this transformation. Thus both bare holon $h_{i}$ and spinon
$a_{i\sigma}$ are not gauge invariant, and they are strongly
coupled by these $U(1)$ gauge field fluctuations. In other words,
these bare holon and spinon are not real.

However, the constrained CP$^{1}$ boson $a_{i\sigma}$ can be
mapped exactly onto the pseudospin representation defined with an
additional phase factor, this is because the empty and doubly
occupied spin states have been ruled out due to the constraint
$a^{\dagger}_{i\uparrow}a_{i\uparrow}+a^{\dagger}_{i\downarrow}
a_{i\downarrow}=1$, and only the spin-up and spin-down singly
occupied spin states are allowed, therefore the original
four-dimensional representation space is reduced to a 2D space.
Due to the symmetry of the spin-up and spin-down states, $\mid
{\rm occupied} \rangle_{\uparrow}=\left (\matrix{1\cr
0\cr}\right)_{\uparrow}$ and $\mid {\rm empty} \rangle_{\uparrow}
=\left (\matrix{0\cr 1\cr}\right)_{\uparrow}$ are singly-occupied
and empty spin-up, while $\mid{\rm occupied}\rangle_{\downarrow}=
\left(\matrix{0\cr 1\cr}\right)_{\downarrow}$ and $\mid{\rm empty}
\rangle_{\downarrow}=\left(\matrix{1\cr 0\cr}\right)_{\downarrow}$
are singly-occupied and empty spin-down states, respectively, thus
the constrained CP$^{1}$ boson operators $a_{i\sigma}$ can be
represented in this reduced 2D space as,
\begin{mathletters}
\begin{eqnarray}
a_{\uparrow}&=&e^{i\Phi_{\uparrow}}\mid {\rm occupied}
\rangle_{\downarrow}~_{\uparrow}\langle {\rm occupied}\mid
\nonumber \\
&=&e^{i\Phi_{\uparrow}}\left (\matrix{0 &0\cr 1 &0\cr}\right)=
e^{i\Phi_{\uparrow}}S^{-}, \\
a_{\downarrow}&=&e^{i\Phi_{\downarrow}}\mid {\rm occupied}
\rangle_{\uparrow}~_{\downarrow}\langle {\rm occupied}\mid
\nonumber \\
&=&e^{i\Phi_{\downarrow}}\left (\matrix{0 &1\cr 0 &0\cr}\right)=
e^{i\Phi_{\downarrow}}S^{+},
\end{eqnarray}
\end{mathletters}
where $S^{-}$ is the $S^{z}$ lowering operator, while $S^{+}$ is
the $S^{z}$ raising operator, then the constraint $\sum_{\sigma}
a^{\dagger}_{i\sigma}a_{i\sigma}=S^{+}_{i}S^{-}_{i}+S^{-}_{i}
S^{+}_{i}=1$ is exactly satisfied. Obviously, the bare spinon
contains both phase and amplitude parts, and the phase part is
described by the phase factor $e^{i\Phi_{i\sigma}}$, while the
amplitude part is described by the spin operator $S_{i}$. In this
case, the electron decoupling form (2) with the constraint can be
expressed as,
\begin{eqnarray}
C_{i\uparrow}=h^{\dagger}_{i}e^{i\Phi_{i\uparrow}}S^{-}_{i},~~~~
C_{i\downarrow}=h^{\dagger}_{i}e^{i\Phi_{i\downarrow}}S^{+}_{i},
\end{eqnarray}
with the local $U(1)$ gauge transformation (4) is rewritten as,
\begin{eqnarray}
h_{i}\rightarrow h_{i}e^{i\theta_{i}}, ~~~
\Phi_{i\sigma}\rightarrow \Phi_{i\sigma}+\theta_{i}.
\end{eqnarray}
Moreover, the phase factor of the bare spinon $e^{i\Phi_{i\sigma}}$
can be incorporated into the bare holon, thus we obtain a new
fermion-spin transformation from Eq. (6) as,
\begin{eqnarray}
C_{i\uparrow}=h^{\dagger}_{i\uparrow}S^{-}_{i},~~~~
C_{i\downarrow}=h^{\dagger}_{i\downarrow}S^{+}_{i},
\end{eqnarray}
with the {\it spinful fermion operator}
$h_{i\sigma}=e^{-i\Phi_{i\sigma}} h_{i}$ describes the charge
degree of freedom together with the phase part of the spin degree
of freedom ({\it dressed holon}), while the spin operator $S_{i}$
describes the amplitude part of the spin degree of freedom ({\it
dressed spinon}). This electron decoupling form (8) is called as a
partial CSS since only the amplitude part of the spin degree of
freedom is separated from the electron operator. These dressed
holon and spinon are invariant under the local $U(1)$ gauge
transformation (7), and therefore all physical quantities from the
dressed holon and spinon also are invariant with respect to this
gauge transformation. In this sense, the dressed holon and spinon
are real. The dressed holon carries a spinon cloud (magnetic
flux), and is a magnetic dressing \cite{dagotto2}. In other words,
the dressed holon carries some spinon messages, i.e., it shares
some effects of spinon configuration rearrangements due to the
presence of the hole itself. We emphasize that the dressed holon
$h_{i\sigma}=e^{-i\Phi_{i\sigma}} h_{i}$ is the spinless fermion
$h_{i}$ (bare holon) incorporated in the spinon cloud
$e^{-i\Phi_{i\sigma}}$ (magnetic flux). Although in the common
sense $h_{i\sigma}$ is not an real spinful fermion operator, it
behaves like a spinful fermion. In correspondence with these
special physical properties, we find that
$h^{\dagger}_{i\sigma}h_{i\sigma}
=h^{\dagger}_{i}e^{i\Phi_{i\sigma}}e^{-i\Phi_{i\sigma}}h_{i}=
h^{\dagger}_{i}h_{i}$, which guarantees that the electron
constraint, $\sum_{\sigma}C^{\dagger}_{i\sigma}C_{i\sigma}=
S^{+}_{i}h_{i\uparrow}h^{\dagger}_{i\uparrow}S^{-}_{i}+S^{-}_{i}
h_{i\downarrow}h^{\dagger}_{i\downarrow}S^{+}_{i}=h_{i}
h^{\dagger}_{i}(S^{+}_{i}S^{-}_{i}+S^{-}_{i}S^{+}_{i})=1-
h^{\dagger}_{i}h_{i}\leq 1$, is always satisfied in analytical
calculations. Moreover the double {\it spinful fermion} occupancy,
$h^{\dagger}_{i\sigma}h^{\dagger}_{i-\sigma}=e^{i\Phi_{i\sigma}}
h^{\dagger}_{i}h^{\dagger}_{i}e^{i\Phi_{i-\sigma}}=0$ and
$h_{i\sigma}h_{i-\sigma}=e^{-i\Phi_{i\sigma}}h_{i}h_{i}
e^{-i\Phi_{i-\sigma}}=0$, is ruled out automatically. Since the
spinless fermion $h_{i}$ and spin operators $S^{+}_{i}$ and
$S^{-}_{i}$ obey the anticommutation relation and Pauli spin
algebra, respectively, it is then easy to show that the spinful
fermion $h_{i\sigma}$ also obeys the same anticommutation relation
as the spinless fermion $h_{i}$.

Although the choice of the CP$^{1}$ representation is convenient,
so long as $h^{\dagger}_{i}h_{i}=1$,
$\sum_{\sigma}C^{\dagger}_{i\sigma} C_{i\sigma}=0$, no matter what
the values of $S^{+}_{i}S^{-}_{i}$ and $S^{-}_{i}S^{+}_{i}$ are,
therefore a "spin" even to an empty site has been assigned. It has
been shown \cite{feng1} that this defect can be cured by
introducing a projection operator $P_{i}$, i.e., the constrained
electron operator can be mapped exactly using the fermion-spin
transformation (8) defined with an additional projection operator
$P_{i}$. However, this projection operator is cumbersome to handle
in the many cases, and it has been dropped in the actual
calculations \cite{feng1,feng2}. It has been shown
\cite{feng1,feng2,plakida} that such treatment leads to errors of
the order $x$ in counting the number of spin states, which is
negligible for small dopings. Moreover, the electron constraint is
still exactly obeyed even in the MF approximation (MFA), and
therefore the essential physics of the gauge invariant dressed
holon and spinon is kept. This is because the constrained electron
operator $C_{i\sigma}$ in the $t$-$J$ type model can also be
mapped onto the slave-fermion formalism \cite{feng2} as
$C_{i\sigma}=h^{\dagger}_{i}b_{i\sigma}$ with the constraint
$h^{\dagger}_{i}h_{i}+\sum_{\sigma}b^{\dagger}_{i\sigma}
b_{i\sigma}=1$. We can solve this constraint by rewriting the
boson operators $b_{i\sigma}$ in terms of the CP$^{1}$ boson
operators $a_{i\sigma}$ as $b_{i\sigma}=a_{i\sigma}
\sqrt{1-h^{\dagger}_{i}h_{i}}$ supplemented by the constraint
$\sum_{\sigma}a^{\dagger}_{i\sigma}a_{i\sigma}=1$. As mentioned
above, the CP$^{1}$ boson operators $a_{i\uparrow}$ and
$a_{i\downarrow}$ with the constraint can be identified with the
pseudospin lowering and raising operators, respectively, defined
with the additional phase factor,  therefore the projection
operator is approximately related to the holon number operator by
$P_{i}\sim\sqrt{1-h^{\dagger}_{i\sigma}h_{i\sigma}}=
\sqrt{1-h^{\dagger}_{i}h_{i}}$, and its main role is to remove the
spurious spin when there is a holon at the site $i$.

\section{Dressed holon and spinon Green's functions}

Before discussing the charge and spin dynamics, let us first
calculate the dressed holon and spinon Green's functions. The
low-energy behavior of the $t$-$t'$-$J$ model in the partial CSS
fermion-spin representation can be expressed as \cite{feng1,feng2},
\begin{eqnarray}
H=&-&t\sum_{i\hat{\eta}}(h_{i\uparrow}S^{+}_{i}
h^{\dagger}_{i+\hat{\eta}\uparrow}S^{-}_{i+\hat{\eta}}+
h_{i\downarrow}S^{-}_{i}h^{\dagger}_{i+\hat{\eta}\downarrow}
S^{+}_{i+\hat{\eta}}) \nonumber \\
&+&t'\sum_{i\hat{\tau}}(h_{i\uparrow}S^{+}_{i}
h^{\dagger}_{i+\hat{\tau}\uparrow}S^{-}_{i+\hat{\tau}}+
h_{i\downarrow}S^{-}_{i}h^{\dagger}_{i+\hat{\tau}\downarrow}
S^{+}_{i+\hat{\tau}}) \nonumber \\
&-&\mu\sum_{i\sigma}h^{\dagger}_{i\sigma}h_{i\sigma}+J_{{\rm eff}}
\sum_{i\hat{\eta}}{\bf S}_{i}\cdot{\bf S}_{i+\hat{\eta}}.
\end{eqnarray}
where $J_{{\rm eff}}=(1-x)^{2}J$, and $x=\langle
h^{\dagger}_{i\sigma}h_{i\sigma}\rangle=\langle h^{\dagger}_{i}
h_{i}\rangle$ is the hole doping concentration. As a consequence,
the kinetic part in the $t$-$t'$-$J$ model has been expressed as
the dressed holon-spinon interaction, which dominates the
essential physics of the underdoped cuprates. The one-particle
dressed holon and spinon two-time Green's functions are defined
as,
\begin{mathletters}
\begin{eqnarray}
g_{\sigma}(i-j,t-t')&=&-i\theta(t-t')\langle [h_{i\sigma}(t),
h^{\dagger}_{j\sigma}(t')]\rangle \nonumber \\
&=&\langle\langle h_{i\sigma}(t);h^{\dagger}_{j\sigma}(t') \rangle
\rangle ,\\
D(i-j,t-t')&=&-i\theta(t-t')\langle [S^{+}_{i}(t),S^{-}_{j}(t')]
\rangle \nonumber \\
&=&\langle\langle S^{+}_{i}(t);S^{-}_{j}(t')\rangle\rangle,
\end{eqnarray}
\end{mathletters}
respectively, where $\langle \ldots \rangle$ is an average over the
ensemble.

\subsection{Equation of motion}

Since the dressed spinon operators obey Pauli algebra, our goal is
to evaluate the dressed holon and spinon Green's functions
directly for the fermion and spin operators in terms of the
equation of motion method. In the framework of the equation of
motion, the time-Fourier transform of the two-time Green's
function $G(\omega)=\langle\langle
A;A^{\dagger}\rangle\rangle_{\omega}$ satisfies the equation
\cite{zubarev}, $\omega\langle\langle A;
A^{\dagger}\rangle\rangle_{\omega}=\langle [A,A^{\dagger}]\rangle+
\langle\langle [A,H];A^{\dagger}\rangle\rangle_{\omega}$. If we
define the orthogonal operator $L$ as, $[A,H]=\zeta A-iL$ with
$\langle [L,A^{\dagger}]\rangle =0$, the full Green's function can
be expressed as,
\begin{eqnarray}
G(\omega)=G^{(0)}(\omega)+{1\over \varsigma^{2}}G^{(0)}(\omega)
\langle\langle L;L^{\dagger}\rangle\rangle_{\omega}G^{(0)}(\omega),
\end{eqnarray}
where $\varsigma=\langle [A,A^{\dagger}]\rangle$, and the MF
Green's function $G^{(0)-1}(\omega)=(\omega-\zeta)/\varsigma$.
It has been shown \cite{zubarev} that if the self-energy
$\Sigma(\omega)$ is identified as the irreducible part of
$\langle\langle L;L^{\dagger}\rangle\rangle_{\omega}$, the full
Green's function (11) can be evaluated as,
\begin{eqnarray}
G(\omega)={\varsigma\over \omega-\zeta-\Sigma(\omega)},
\end{eqnarray}
with $\Sigma(\omega)=\langle\langle L;L^{\dagger}\rangle
\rangle ^{irr}_{\omega}/\varsigma$. In the framework of the
diagrammatic technique, $\Sigma(\omega)$ corresponds to the
contribution of irreducible diagrams.

\subsection{The mean-field theory}

Within MFA, the $t$-$t'$-$J$ model (9) can be decoupled as,
\begin{mathletters}
\begin{eqnarray}
H_{MFA}&=&H_{t}+H_{J}-8Nt\chi_{1}\phi_{1}+8Nt'\chi_{2}\phi_{2},\\
H_{t}&=&\chi_{1}t\sum_{i\hat{\eta}\sigma}
h^{\dagger}_{i+\hat{\eta}\sigma}h_{i\sigma}-\chi_{2}
t'\sum_{i\hat{\tau}\sigma}h^{\dagger}_{i+\hat{\tau}\sigma}
h_{i\sigma} \nonumber \\
&-&\mu\sum_{i\sigma}h^{\dagger}_{i\sigma}h_{i\sigma}, \\
H_{J}&=&J_{{\rm eff}}\sum_{i\hat{\eta}}[{1\over 2}\epsilon
(S^{+}_{i}S^{-}_{i+\hat{\eta}}+S^{-}_{i}S^{+}_{i+\hat{\eta}})+
S^{Z}_{i}S^{Z}_{i+\hat{\eta}}]\nonumber \\
&-&\phi_{2}t'\sum_{i\hat{\tau}}
(S^{+}_{i}S^{-}_{i+\hat{\tau}}+S^{-}_{i}S^{+}_{i+\hat{\tau}}),
\end{eqnarray}
\end{mathletters}
where the dressed holon's particle-hole parameters $\phi_{1}=
\langle h^{\dagger}_{i\sigma}h_{i+\hat{\eta}\sigma}\rangle$ and
$\phi_{2}=\langle h^{\dagger}_{i\sigma}h_{i+\hat{\tau}\sigma}
\rangle$, the dressed spinon correlation functions
$\chi_{1}=\langle S_{i}^{+}S_{i+\hat{\eta}}^{-}\rangle$ and
$\chi_{2}=\langle S_{i}^{+}S_{i+\hat{\tau}}^{-}\rangle$, and
$\epsilon=1+2t\phi_{1}/J_{{\rm eff}}$. Since AFLRO in the undoped
cuprates is destroyed \cite{kbse6} by hole doping of the order
$\sim 0.024$, there is therefore no AFLRO in the doped regime
$x\geq 0.025$, i.e., $\langle S^{z}_{i}\rangle =0$, and a
disordered spin liquid state emerges. It has been argued that this
spin liquid state may play a crucial role in the mechanism for
HTSC \cite{pwa1,pwa2}. In this paper, we focus on the normal-state
properties in the doped regime without AFLRO. In this case, a
similar MF theory \cite{feng3} of the $t$-$J$ model based on the
fermion-spin theory has been discussed within the Kondo-Yamaji
decoupling scheme \cite{kondo}, which is a stage one-step further
than the Tyablikov's decoupling scheme \cite{tyablikov}. In this
MF theory \cite{feng3}, the phase factor $e^{i\Phi_{i\sigma}}$
describing the phase part of the spin degree of freedom was not
considered. Following their discussions \cite{feng3}, we obtain
the MF dressed holon and spinon Green's functions in the present
case as,
\begin{mathletters}
\begin{eqnarray}
g^{(0)}_{\sigma}({\bf k},\omega)&=&{1\over \omega-\xi_{k}}, \\
D^{(0)}({\bf k},\omega)&=&{B_{k}\over\omega^{2}-\omega_{k}^{2}},
\end{eqnarray}
\end{mathletters}
respectively, where $B_{k}=\lambda_{1}[2\chi^{z}_{1}(\epsilon
\gamma_{{\bf k}}-1)+\chi_{1}(\gamma_{{\bf k}}-\epsilon)]-
\lambda_{2}(2\chi^{z}_{2}\gamma'_{{\bf k}}-\chi_{2})$,
$\lambda_{1}=2ZJ_{eff}$, $\lambda_{2}=4Z\phi_{2}t'$,
$\gamma_{{\bf k}}=(1/Z)\sum_{\hat{\eta}}e^{i{\bf k}\cdot
\hat{\eta}}$, $\gamma'_{{\bf k}}=(1/Z)\sum_{\hat{\tau}}
e^{i{\bf k}\cdot\hat{\tau}}$, $Z$ is the number of the nearest
neighbor or second-nearest neighbor sites, and the MF dressed holon
and spinon excitation spectra are given by,
\begin{mathletters}
\begin{eqnarray}
\xi_{k}&=&Zt\chi_{1}\gamma_{{\bf k}}-Zt'\chi_{2}
\gamma'_{{\bf k}}-\mu, \\
\omega^{2}_{k}&=&A_{1}(\gamma_{k})^{2}+A_{2}(\gamma'_{k})^{2}+
A_{3}\gamma_{k}\gamma'_{k} \nonumber \\
&+&A_{4}\gamma_{k}+A_{5}\gamma'_{k}+A_{6},
\end{eqnarray}
\end{mathletters}
respectively, with $A_{1}=\alpha\epsilon\lambda_{1}^{2}(\epsilon
\chi^{z}_{1}+\chi_{1}/2)$, $A_{2}=\alpha\lambda_{2}^{2}
\chi^{z}_{2}$, $A_{3}=-\alpha\lambda_{1}\lambda_{2}(\epsilon
\chi^{z}_{1}+\epsilon\chi^{z}_{2}+\chi_{1}/2)$, $A_{4}=-\epsilon
\lambda_{1}^{2}[\alpha(\chi^{z}_{1}+\epsilon\chi_{1}/2)+(\alpha
C^{z}_{1}+(1-\alpha)/(4Z)-\alpha\epsilon\chi_{1}/(2Z))+(\alpha
C_{1}+(1-\alpha)/(2Z)-\alpha\chi^{z}_{1}/2)/2]+\alpha\lambda_{1}
\lambda_{2}(C_{3}+\epsilon\chi_{2})/2$, $A_{5}=-3\alpha
\lambda^{2}_{2}\chi_{2}/(2Z)+\alpha\lambda_{1}\lambda_{2}
(\chi^{z}_{1}+\epsilon\chi_{1}/2+C^{z}_{3})$, $A_{6}=
\lambda^{2}_{1}[\alpha C^{z}_{1}+(1-\alpha)/(4Z)-\alpha\epsilon
\chi_{1}/(2Z)+\epsilon^{2}(\alpha C_{1}+(1-\alpha)/(2Z)-\alpha
\chi^{z}_{1}/2)/2]+\lambda^{2}_{2}(\alpha C_{2}+(1-\alpha)/(2Z)-
\alpha\chi^{z}_{2}/2)/2)-\alpha\epsilon\lambda_{1}\lambda_{2}
C_{3}$, and the spinon correlation functions
$\chi^{z}_{1}=\langle S_{i}^{z}S_{i+\hat{\eta}}^{z}\rangle$,
$\chi^{z}_{2}=\langle S_{i}^{z}S_{i+\hat{\tau}}^{z}\rangle$,
$C_{1}=(1/Z^{2})\sum_{\hat{\eta},\hat{\eta'}}\langle
S_{i+\hat{\eta}}^{+}S_{i+\hat{\eta'}}^{-}\rangle$,
$C^{z}_{1}=(1/Z^{2})\sum_{\hat{\eta},\hat{\eta'}}\langle
S_{i+\hat{\eta}}^{z}S_{i+\hat{\eta'}}^{z}\rangle$,
$C_{2}=(1/Z^{2})\sum_{\hat{\tau},\hat{\tau'}}\langle
S_{i+\hat{\tau}}^{+}S_{i+\hat{\tau'}}^{-}\rangle$,
$C_{3}=(1/Z)\sum_{\hat{\tau}}\langle S_{i+\hat{\eta}}^{+}
S_{i+\hat{\tau}}^{-}\rangle$, and $C^{z}_{3}=(1/Z)
\sum_{\hat{\tau}}\langle S_{i+\hat{\eta}}^{z}
S_{i+\hat{\tau}}^{z}\rangle$.
In order not to violate the sum rule of the correlation function
$\langle S^{+}_{i}S^{-}_{i}\rangle=1/2$ in the case without
AFLRO, the important decoupling parameter $\alpha$ has been
introduced in the MF calculation \cite{feng3,kondo}, which can be
regarded as the vertex correction. All the above MF order
parameters, decoupling parameter $\alpha$, and chemical potential
$\mu$ are determined by the self-consistent calculation
\cite{feng3}.

\subsection{The dressed holon and spinon self-energies}

With the help of Eq. (12), the full dressed holon and spinon
Green's functions of the $t$-$t'$-$J$ model (9) are expressed as,
\begin{mathletters}
\begin{eqnarray}
g_{\sigma}({\bf k},\omega)&=&{1\over \omega-\xi_{k}-
\Sigma^{(2)}_{h}({\bf k},\omega)}, \\
D({\bf k},\omega)&=&{B_{k}\over \omega^{2} -\omega^{2}_{k}-
\Sigma^{(2)}_{s}({\bf k},\omega)},
\end{eqnarray}
\end{mathletters}
respectively, where the second-order dressed holon self-energy from
the dressed spinon pair bubble $\Sigma^{(2)}_{h}({\bf k},\omega)=
\langle\langle L^{(h)}_{k}(t);L^{(h)\dagger}_{k}(t')\rangle
\rangle_{\omega}$ with the orthogonal operator $L^{(h)}_{i}=-t
\sum_{\hat{\eta}}h_{i+\hat{\eta}\sigma}(S^{-}_{i+\hat{\eta}}
S^{+}_{i}-\chi_{1})+t'\sum_{\hat{\tau}}h_{i+\hat{\tau}\sigma}(
S^{-}_{i+\hat{\tau}}S^{+}_{i}-\chi_{2})$, and can be evaluated as
\cite{feng2},
\begin{eqnarray}
\Sigma_{h}^{(2)}({\bf k},\omega)&=&{1\over 2}\left ({Z\over N}
\right )^2\sum_{pp'}\gamma^{2}_{12}({\bf k,p,p'}){B_{p'}B_{p+p'}
\over 4\omega_{p'}\omega_{p+p'}}\nonumber \\
&\times& \left ({F^{(h)}_{1}(k,p,p')\over\omega+\omega_{p+p'}-
\omega_{p'}-\xi_{p+k}}\right. \nonumber \\
&+& {F^{(h)}_{2}(k,p,p')\over\omega+\omega_{p'}-\omega_{p+p'}
-\xi_{p+k}} \nonumber \\
&+& {F^{(h)}_{3}(k,p,p')\over\omega+\omega_{p'}+
\omega_{p+p'}-\xi_{p+k}} \nonumber \\
&-&\left . {F^{(h)}_{4}(k,p,p')\over\omega -
\omega_{p+p'}-\omega_{p'}-\xi_{p+k}}\right ),
\end{eqnarray}
where $\gamma^{2}_{12}({\bf k,p,p'})=[(t\gamma_{{\bf p'+p+k}}-t'
\gamma'_{{\bf p'+p+k}})^{2}+(t\gamma_{{\bf p'-k}}-t' \gamma'_{{\bf
p'-k}})^{2}]$, $F^{(h)}_{1}(k,p,p')=n_{F}(\xi_{p+k})
[n_{B}(\omega_{p'})-n_{B}(\omega_{p+p'})]+n_{B}(\omega_{p+p'})[1+
n_{B}(\omega_{p'})]$, $F^{(h)}_{2}(k,p,p')=n_{F}(\xi_{p+k})[n_{B}
(\omega_{p'+p})-n_{B}(\omega_{p'})]+n_{B}(\omega_{p'})[1+n_{B}(
\omega_{p'+p})]$, $F^{(h)}_{3}(k,p,p')=n_{F}(\xi_{p+k})[1+n_{B}
(\omega_{p+p'})+n_{B}(\omega_{p'})]+n_{B}(\omega_{p'})n_{B}
(\omega_{p+p'})$, $F^{(h)}_{4}(k,p,p')=n_{F}(\xi_{p+k)}[1+n_{B}
(\omega_{p+p'})+n_{B}(\omega_{p'})]-[1+n_{B}(\omega_{p'})][1+n_{B}
(\omega_{p+p'})]$, and $n_{B}(\omega_{p})$ and $n_{F}(\xi_{p})$
are the boson and fermion distribution functions, respectively.
This dressed holon self-energy is ascribed purely to the dressed
holon-spinon interaction, and characterizes the competition
between the kinetic energy and magnetic energy. The calculation of
the dressed spinon self-energy is quite tedious \cite{feng4},
since our starting point is the dressed spinon MF solution
\cite{feng3} within the Kondo-Yamaji decoupling scheme in
subsection B. From Eq. (11) and the MF dressed spinon Green's
function (14b), the full dressed spinon Green's function satisfies
the relation \cite{feng4},
$\omega^{2}D(k,\omega)=B_{k}+\langle\langle
[[S^{+}_{i}(t),H(t)],H(t)];S^{-}_{j}(t')\rangle\rangle_{k,\omega}$
with
$[[S^{+}_{i},H],H]_{k}=\omega^{2}_{k}S^{+}_{k}-i\Gamma^{(s)}_{k}$.
In the disordered liquid state without AFLRO, the dressed
holon-spinon interaction should dominate the essential physics
\cite{feng4}. In this case, the orthogonal operator $L^{(s)}_{k}$
for the dressed spinon can be selected from $\Gamma^{(s)}_{k}$ as
\cite{feng4},
\begin{eqnarray}
L^{(s)}_{i}&=&-(2\epsilon\chi^{z}_{1}+\chi_{1})\lambda_{1}
{1\over Z}\sum_{\hat{\eta},\hat{a}}t_{\hat{a}}
(h^{\dagger}_{i+\hat{\eta}\uparrow}h_{i+\hat{\eta}+\hat{a}\uparrow}
\nonumber \\
&+& h^{\dagger}_{i+\hat{\eta}+\hat{a}\downarrow}
h_{i+\hat{\eta}\downarrow}-2\phi_{\hat{a}})
S^{+}_{i+\hat{\eta}+\hat{a}} \nonumber \\
&+&[(2\chi^{z}_{1}+\epsilon\chi_{1})\lambda_{1}-\chi_{2}
\lambda_{2}]\sum_{\hat{a}}t_{\hat{a}}(h^{\dagger}_{i\uparrow}
h_{i+\hat{a}\uparrow} \nonumber \\
&+&h^{\dagger}_{i+\hat{a}\downarrow}h_{i\downarrow}-
2\phi_{\hat{a}})S^{+}_{i+\hat{a}} \nonumber \\
&+&2\chi^{z}_{2}\lambda_{2}{1\over Z}\sum_{\hat{\tau},\hat{a}}
t_{\hat{a}}(h^{\dagger}_{i+\hat{\tau}\uparrow}
h_{i+\hat{\tau}+\hat{a}\uparrow} \nonumber \\
&+& h^{\dagger}_{i+\hat{\tau}+\hat{a}\downarrow}
h_{i+\hat{\tau}\downarrow}-2\phi_{\hat{a}})
S^{+}_{i+\hat{\tau}+\hat{a}},
\end{eqnarray}
where $\hat{a}=\hat{\eta},\hat{\tau}$, with $t_{\hat{\eta}}=t$,
$\phi_{\hat{\eta}}=\phi_{1}$, and $t_{\hat{\tau}}=-t'$,
$\phi_{\hat{\tau}}=\phi_{2}$. Following [38], we obtain the
dressed spinon self-energy $\Sigma^{(2)}_{s}({\bf k},\omega)=
\langle\langle L^{(s)}_{i}(t);L^{(s)\dagger}_{j}(t')\rangle
\rangle_{k,\omega}$,
\begin{eqnarray}
\Sigma_{s}^{(2)}({\bf k},\omega)&=&B_{k}\left ({Z\over N}
\right )^{2}\sum_{pp'}\gamma^{2}_{12}({\bf k,p,p'})
{B_{k+p}\over 2\omega_{k+p}} \nonumber \\
&\times&\left ({F^{(s)}_{1}(k,p,p')\over\omega+\xi_{p+p'}-\xi_{p'}-
\omega_{k+p}} \right .\nonumber \\
&-&\left . {F^{(s)}_{2}(k,p,p')\over \omega+\xi_{p+p'}-
\xi_{p'}+\omega_{k+p}}\right ),
\end{eqnarray}
with $F^{(s)}_{1}(k,p,p')=n_{F}(\xi_{p+p'})[1-n_{F}(\xi_{p'})]-
n_{B}(\omega_{k+p})[n_{F}(\xi_{p'})-n_{F}(\xi_{p+p'})]$, and
$F^{(s)}_{2}(k,p,p')=n_{F}(\xi_{p+p'})[1-n_{F}(\xi_{p'})]+[1+n_{B}
(\omega_{k+p})][n_{F}(\xi_{p'})-n_{F}(\xi_{p+p'})]$. Within the
diagrammatic technique, this dressed spinon self-energy
$\Sigma_{s}^{(2)}({\bf k},\omega)$ corresponds to the contribution
from the dressed holon pair bubble, and is consistent with our
previous result \cite{feng4}.

\section{Charge transport}

Recently, the emergence and evolution of the metallic transport
in doped cuprates have been extensively studied by virtue of
systematic transport measurements \cite{ando1}. It is shown that
the resistivity shows a crossover from the low temperature
insulating-like to moderate temperature metallic-like behavior in
the heavily underdoped regime ($0.025\leq x<0.055$), and a
temperature linear dependence with deviations at low temperatures
in the underdoped regime ($0.055< x<0.15$). These striking
behaviors have been found to be intriguingly related to the AF
correlation \cite{ando1}. In this case, a natural question is
what is the physical origin of this transport transformation from
the insulating liquid state in the heavily underdoped regime to the
unusual metallic state in the underdoped regime? In this section,
we try to discuss this issue. Since the local constraint has been
treated properly in the partial CSS fermion-spin theory, the extra
$U(1)$ gauge degree of freedom related with the local constraint
is incorporated into the dressed holon as mentioned in Sec. II. In
this case, the external electronic field only is coupled to the
dressed holons, and the conductivity is given by,
\begin{eqnarray}
\sigma(\omega)=-{{\rm Im}\Pi_{h}(\omega)\over\omega},
\end{eqnarray}
with $\Pi_{h}(\omega)$ is the dressed holon current-current
correlation function, and is defined as $\Pi_{h}(t-t')=\langle
\langle j_{h}(t)j_{h}(t')\rangle\rangle$, where the current density
of the dressed holons is obtained by taking the time derivation of
the polarization operator with the use of the equation of motion as
\cite{feng2}, $j_{h}=(e\chi_{1}t/2)\sum_{i\hat{\eta}\sigma}
\hat{\eta}h_{i+\hat{\eta}\sigma}^{\dagger}h_{i\sigma}-
(e\chi_{2}t'/2)\sum_{i\hat{\tau}\sigma}\hat{\tau}
h_{i+\hat{\tau}\sigma}^{\dagger}h_{i\sigma}$. With the help of the
full dressed holon Green's function (16a), the current-current
correlation function is evaluated as,
\begin{eqnarray}
\Pi_{h}(i\omega_{n})&=&-({Ze\over 2})^{2}{1\over N}\sum_{k}
\gamma_{s}^{2}({\bf k}){1\over \beta}\sum_{i\omega_{m}'\sigma}
g_{\sigma}({\bf k},i\omega_{m}'+i\omega_{n}) \nonumber \\
&\times& g_{\sigma}({\bf k},i\omega_{m}'),
\end{eqnarray}
where $i\omega_{n}$ is the Matsubara frequency, $\gamma^{2}_{s}
({\bf k})=[\sin^{2} k_{x}(\chi_{1}t-2\chi_{2}t'\cos k_{y})^{2}+
\sin^{2} k_{y}(\chi_{1}t-2\chi_{2}t'\cos k_{x})^{2}]/4$. The full
dressed holon Green's function can be expressed as frequency
integrals in terms of the spectral representation,
\begin{eqnarray}
g_{\sigma}({\bf k},i\omega_{n})=\int_{-\infty}^{\infty}
{d\omega\over 2\pi}{A^{(h)}_{\sigma}({\bf k},\omega)\over
i\omega_{n}-\omega},
\end{eqnarray}
with the dressed holon spectral function $A^{(h)}_{\sigma}({\bf k},
\omega)=-2{\rm Im}g_{\sigma}({\bf k},\omega)$. Substituting the
Eqs. (22) and (21) into Eq. (20), and evaluating the frequency
summation, we obtain the conductivity as \cite{feng2},
\begin{eqnarray}
\sigma(\omega)&=&({Ze\over 2})^2{1\over N}\sum_{k\sigma}
\gamma_{s}^{2}({\bf k})\int^{\infty}_{-\infty}{d\omega'\over 2\pi}
A^{(h)}_{\sigma}({\bf k},\omega'+\omega) \nonumber \\
&\times& A^{(h)}_{\sigma}({\bf k},\omega'){n_{F}(\omega'+\omega)-
n_{F}(\omega') \over \omega}.
\end{eqnarray}
We have performed a numerical calculation for $\sigma(\omega)$, and
the results at $x=0.03$ (solid line), $x=0.05$ (dashed line), and
$x=0.07$ (dotted line) for $t/J=2.5$ and $t'/t=0.15$ with $T=0$ are
plotted in Fig. 1, hereinafter the charge $e$ is set as the unit.
For a comparison, the experimental result \cite{cooper1} of
YBa$_{2}$Cu$_{3}$O$_{7-x}$ is also plotted in Fig. 1 (inset). Our
results show that there is a low-energy peak at $\omega< 0.3t$
separated by a gap or pseudogap at $0.3t$ from a midinfrared band.
This midinfrared band is doping dependent, the component increases
with increasing dopings for $0.3t<\omega <1.0t$ and is nearly
independent of dopings for $\omega >1.0t$, however, the position of
the midinfrared peak is shifted to lower energies with increased
dopings. This reflects an increase in the mobile carrier density,
and indicates that the spectral weight of the midinfrared sideband
is taken from the Drude absorption, then the spectral weight from
both low energy peak and midinfrared band represent the actual
free-carrier density. For a better understanding of the optical
properties, we have made a series of calculations for
$\sigma (\omega)$ at different temperatures, and the results at
$x=0.06$ with $T=0$ (solid line), $T=0.1J$ (dashed line), and
$T=0.3J$ (dotted line) for $t/J=2.5$ and $t'/t=0.15$ are plotted
in Fig. 2 in comparison with the experimental data \cite{cooper1}
taken from YBa$_{2}$Cu$_{3}$O$_{7-x}$ (inset). It is shown that
$\sigma(\omega)$ is temperature-dependent for $\omega <1.0t$, and
almost temperature-independent for $\omega>1.0t$. The peak at low
energies broadens and decreases in the height with increasing
temperatures, while the component in the low energy region
increases with increasing temperatures, then there is a tendency
towards the Drude-like behavior. The midinfrared band is severely
suppressed with increasing temperatures, and vanishes at high
temperatures, in qualitative agreement with experiments
\cite{cooper1,uchida1}.

\begin{figure}[prb]
\epsfxsize=2.0in\centerline{\epsffile{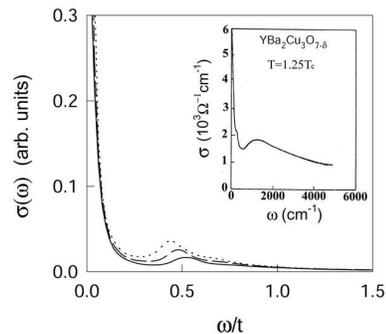}} \caption{The
conductivity at $x=0.03$ (solid line), $x=0.05$ (dashed line), and
$x=0.07$ (dotted line) with $t/J=2.5$ and $t'/t=0.15$ in $T=0$.
Inset: the experimental result of YBa$_{2}$Cu$_{3}$O$_{7-x}$ taken
from Ref. [11].}
\end{figure}
\begin{figure}[prb]
\epsfxsize=2.0in\centerline{\epsffile{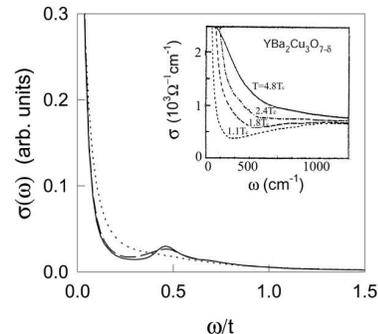}} \caption{The
conductivity at $x=0.06$ in $T=0$ (solid line), $T=0.1J$ (dashed
line), and $T=0.3J$ (dotted line) with $t/J=2.5$ and $t'/t=0.15$.
Inset: the experimental result of YBa$_{2}$Cu$_{3}$O$_{7-x}$ taken
from Ref. [11].}
\end{figure}

Now we turn to discuss the resistivity, which is evaluated as,
$\rho=1/\sigma_{dc}$, with the $dc$ conductivity $\sigma_{dc}$ is
obtained from Eq. (23) as $\sigma_{dc}=\lim_{\omega\rightarrow 0}
\sigma(\omega)$. This resistivity has been evaluated numerically,
and the results are plotted in Fig. 3 as a function of temperature
at $x=0.03$ (solid line), $x=0.04$ (dashed line), $x=0.05$ (dotted
line), and $x=0.06$ (dash-dotted line) for $t/J=2.5$ and
$t'/t=0.15$ in comparison with the experimental data \cite{ando1}
taken from La$_{2-x}$Sr$_{x}$CuO$_{4}$ (inset). Our results show
obviously that the resistivity is characterized by a crossover
from the moderate temperature metallic-like to low temperature
insulating-like behavior in the heavily underdoped regime, and a
temperature linear dependence with deviations at low temperatures
in the underdoped regime. But even in the heavily underdoped
regime, the resistivity exhibits the metallic-like behavior over a
wide range of temperatures, which also is in qualitative agreement
with experiments \cite{ando1}.

\begin{figure}[prb]
\epsfxsize=2.0in\centerline{\epsffile{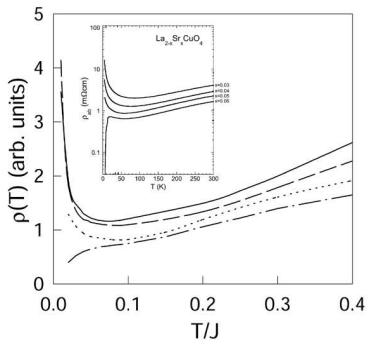}} \caption{The
electron resistivity as a function of temperature at $x=0.03$
(solid line), $x=0.04$ (dashed line), $x=0.05$ (dotted line), and
$x=0.06$ (dash-dotted line) with $t/J=2.5$ and $t'/t=0.15$. Inset:
the experimental result of La$_{2-x}$Sr$_{x}$CuO$_{4}$ taken from
Ref. [14].}
\end{figure}

The perovskite parent compound of doped cuprates is a Mott
insulator, when holes are doped into this insulator, there is a
gain in the kinetic energy per hole proportional to $t$ due to
hopping, but at the same time, the spin correlation is destroyed,
costing an energy of approximately $J$ per site. Thus doped holes
in a Mott insulator can be considered as a competition between the
kinetic energy ($xt$) and magnetic energy ($J$). The magnetic
energy $J$ favors the magnetic order for spins and results in
frustration of the kinetic energy, while the kinetic energy $xt$
favors delocalization of holes and tends to destroy the magnetic
order. In the present partial CSS fermion-spin theory, the
scattering of dressed holons dominates the charge transport, since
the scattering rate is obtained from the dressed holon self-energy
$\Sigma_{h}^{(2)}({\bf k},\omega)$, while this self-energy is
evaluated by considering the dressed holon-spinon interaction, and
characterizes a competition between the kinetic energy and
magnetic energy. In this case, the striking behavior in the
resistivity is intriguingly related to this competition. In the
heavily underdoped regime, the dressed holon kinetic energy is
much smaller than the dressed spinon magnetic energy at lower
temperatures due to the strong AF correlation, where the dressed
holons are localized, and the scattering rate from the dressed
holon self-energy is severely reduced, this leads to the
insulating-like behavior in the resistivity. With increasing
temperatures, the dressed holon kinetic energy is increased, while
the dressed spinon magnetic energy is decreased. In the region
where the dressed holon kinetic energy is larger than the dressed
spinon magnetic energy at moderate temperatures, the dressed
holons can move in the background of the dressed spinon
fluctuation, then the dressed holon scattering would give rise to
the metallic-like behavior in the resistivity. Since the charge
transport is governed by the dressed holon scattering, then $x$
dressed holons are responsible for the electron conductivity.

\section{Incommensurate spin dynamics}

The interplay between AF correlation and HTSC in doped cuprates is
now well-established \cite{kbse1}, but its full understanding is
still a challenging issue. Experimentally, NMR, NQR, and $\mu SR$
techniques, particularly inelastic neutron scattering, can provide
rather detailed information on the spin dynamics of doped cuprates
\cite{rossat,kbse2,kbse3,dai1,kbse4,kbse5}. It has been shown
\cite{kbse3,dai1} that when AFLRO is suppressed, the IC magnetic
correlation develops at a quartet of wave vector
$[\pi(1\pm\delta),\pi]$ and $[\pi,(1\pm\delta)\pi]$, where the
incommensurability $\delta$ increases almost linearly with the
hole doping concentration $x$ at lower dopings, and saturates at
higher dopings. These exotic features are fully confirmed by the
data both on the normal and superconducting states
\cite{kbse3,dai1}. It has been argued that the emergence of the IC
magnetic correlation is due to dopings \cite{kbse7}. Although a
sharp resonance peak at the commensurate AF wave vector has been
observed in some optimally doped samples, this commensurate
scattering is the main new feature that appears in the
superconducting phase \cite{mook1,dai1}. In this section, we only
discuss the IC magnetic correlation in the normal-state. Within
the present partial CSS fermion-spin theory, the spin fluctuation
couples only to the dressed spinons, then the dynamical spin
structure factor (DSSF) is obtained in terms of the full dressed
spinon Green's function (16b) as \cite{feng4},
\begin{eqnarray}
S({\bf k},\omega)&=&-2[1+n_{B}(\omega)]{\rm Im}D(k,\omega)
\nonumber \\
&=&-2[1+n_{B}(\omega)] \nonumber \\
&\times& {B_{k}{\rm Im}\Sigma_{s}^{(2)}({\bf k},\omega)\over
[\omega^{2}-\omega^{2}_{k}-{\rm Re}\Sigma_{s}^{(2)}
({\bf k},\omega)]^{2}+[{\rm Im}\Sigma_{s}^{(2)}
({\bf k},\omega)]^{2}},
\end{eqnarray}
where ${\rm Im}\Sigma_{s}^{(2)}({\bf k},\omega)$ and ${\rm Re}
\Sigma_{s}^{(2)}({\bf k},\omega)$ are the corresponding imaginary
part and real part of the dressed spinon self-energy function
$\Sigma_{s}^{(2)}({\bf k},\omega)$ in Eq. (19).

\begin{figure}[prb]
\epsfxsize=1.7in\centerline{\epsffile{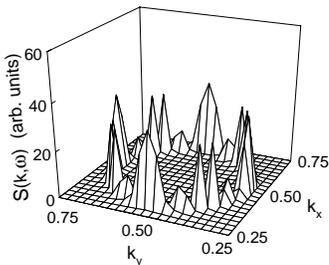}} \caption{The
dynamical spin structure factor spectrum in the $(k_{x},k_{y})$
plane at $x=0.06$ in $T=0.05J$ and $\omega=0.05J$ with $t/J=2.5$
and $t'/t=0.15$.}
\end{figure}
\begin{figure}[prb]
\epsfxsize=1.7in\centerline{\epsffile{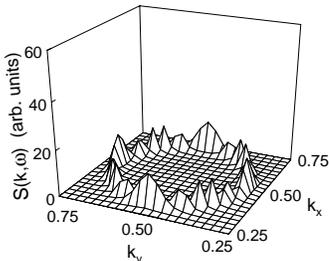}} \caption{The
dynamical spin structure factor spectrum in the $(k_{x},k_{y})$
plane at $x=0.06$ with $t/J=2.5$ and $t'/t=0.15$ in $T=0.05J$ and
$\omega=0.1J$.}
\end{figure}

At the half-filling, the spin fluctuation scattering remains
commensurate at the AF wave vector ${\bf Q}=[1/2,1/2]$ position
(hereafter we use units of $[2\pi,2\pi]$), which is not presented
here for the sake of space. Instead, we plot the DSSF spectrum
$S({\bf k},\omega)$ in the ($k_{x},k_{y}$) plane at $x=0.06$ with
$T=0.05J$ and $\omega=0.05J$ for $t/J=2.5$ and $t'/t=0.15$ in Fig.
4. This result shows that with dopings, there is a commensurate-IC
transition in the spin fluctuation geometry, where all IC peaks
lie on a circle of radius of $\delta$. Although some IC satellite
diagonal peaks appear, the main weight of the IC peaks is in the
parallel direction, and these parallel peaks are located at
$[(1\pm\delta)/2,1/2]$ and $[1/2,(1\pm \delta)/2]$. The IC peaks
are very sharp at low temperatures and energies, which means that
these low energy excitations have a dynamical coherence length at
low temperatures that is larger than the instantaneous correlation
length. For considering IC magnetic fluctuation at a relatively
high energy, we have made a series of scans for $S({\bf
k},\omega)$ with several energies, and the result at $x=0.06$ in
$t/J=2.5$ and $t'/t=0.15$ with $T=0.05J$ for $\omega=0.1J$ is
shown in Fig. 5. Comparing it with Fig. 4 for the same set of
parameters except for $\omega=0.05J$, we see that at low
temperatures, although the positions of the IC peaks are energy
independent, the IC peaks broaden and weaken in amplitude as the
energy increase, and vanish at high energies. This reflects that
the excitation width increases with increasing energies, and thus
leads to that the lifetime of the excitations decreasing quickly
with increasing energies. The present DSSF spectrum has been used
to extract the doping dependence of the incommensurability
$\delta(x)$, which is defined as the deviation of the peak
position from the AF wave vector position, and the result is
plotted in Fig. 6 in comparison with the experimental result
\cite{kbse3} taken from La$_{2-x}$Sr$_{x}$CuO$_{4}$ (inset). Our
result shows that $\delta(x)$ increases progressively with the
doping concentration at lower dopings, but saturates at higher
dopings, in qualitative agreement with experiments
\cite{kbse3,dai1}.

\begin{figure}[prb]
\epsfxsize=2.0in\centerline{\epsffile{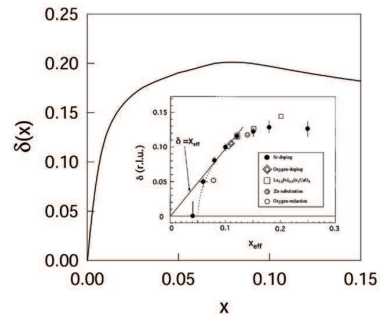}} \caption{The doping
dependence of the incommensurability $\delta(x)$. Inset: the
experimental result of La$_{2-x}$Sr$_{x}$CuO$_{4}$ taken from Ref.
[7].}
\end{figure}

\begin{figure}[prb]
\epsfxsize=2.0in\centerline{\epsffile{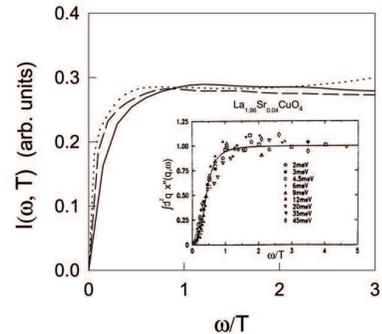}} \caption{The
integrated dynamical spin susceptibility at $x=0.12$ with
$t/J=2.5$ and $t'/t=0.15$ in $T=0.2J$ (solid line), $T=0.3J$
(dashed line), and $T=0.5J$ (dotted line). Inset: the experimental
result of La$_{2-x}$Sr$_{x}$CuO$_{4}$ taken from Ref. [9].}
\end{figure}

The universal integrated dynamical spin response is a
characteristic feature, and is closely related to many other
normal-state properties \cite{kbse1}. The integrated dynamical
spin response is manifested by the integrated dynamical spin
susceptibility (IDSS), and is expressed as,
\begin{eqnarray}
I(\omega, T)={1\over N}\sum_{k}\chi^{\prime\prime}
({\bf k},\omega),
\end{eqnarray}
where the dynamical spin susceptibility is related to DSSF by the
fluctuation dissipation theorem as, $\chi''({\bf k},\omega)=
(1-e^{-\beta\omega})S({\bf k},\omega)=-2{\rm Im}D({\bf k},\omega)$.
This IDSS has been evaluated numerically, and the results at
$x=0.12$ for $t/J=2.5$ and $t'/t=0.15$ with $T=0.2J$ (solid line),
$T=0.3J$ (dashed line), and $T=0.5J$ (dotted line) are plotted in
Fig. 7 in comparison with the experimental data \cite{kbse4} taken
from La$_{2-x}$Sr$_{x}$CuO$_{4}$ (inset). It is shown that the
shape of IDSS appears to be particularly universal, and can be
scaled approximately as $I(\omega,T)\propto {\rm arctan}
[a_{1}\omega/T+a_{3}(\omega/T)^{3}]$, where $I(\omega, T)$ is
almost constant for $\omega/T>1$, and then begin to decrease with
decreasing $\omega/T$ for $\omega/T<1$, also in qualitative
agreement with experiments \cite{kbse4,kbse5}.

Although the scattering of the dressed spinons dominates the spin
dynamics, the effect of the dressed holons on the dressed spinon
part is critical in determining the characteristic feature of the
IC magnetic correlation, which can be understood from the
properties of the dressed spinon excitation spectrum
$E^{2}_{k}=\omega^{2}_{k}+{\rm Re}\Sigma^{(2)}_{s}({\bf
k},E_{k})$. During the calculation of DSSF spectrum in Eq. (24),
we find when $W({\bf
k}_{\delta},\omega)=[\omega^{2}-\omega^{2}_{k_{\delta}}- {\rm
Re}\Sigma_{s}^{(2)}({\bf k}_{\delta},\omega)]^{2}\sim 0$ at some
critical wave vectors $\pm{\bf k}_{\delta}$ in low energies, the
IC peaks appear, then the weight of the IC peaks is dominated by
the inverse of the imaginary part of the dressed spinon
self-energy $1/{\rm Im}\Sigma_{s}^{(2)}({\bf k}_{\delta},\omega)$.
Thus the positions of the IC peaks are determined by both
functions $W({\bf k},\omega)$ and ${\rm Im}\Sigma_{s}^{(2)}({\bf
k},\omega)$, where the zero points of $W({\bf k},\omega)$ (then
the critical wave vectors ${\bf k}_{\delta}$) is doping
dependence. Near the half-filling, the zero point of $W({\bf
k},\omega)$ locates at the AF wave vector [$1/2,1/2$], so the
commensurate AF peak appears there. With dopings, the holes
disturb the AF background. Within the partial CSS framework, as a
result of the self-consistent motion of the dressed holons and
spinons, the IC magnetic correlation is developed beyond a certain
critical doping, this reflects the fact that the low energy spin
excitations drift away from the AF wave vector, or the zero point
of $W({\bf k}_{\delta},\omega)$ is shifted from $[1/2,1/2]$ to
${\bf k}_{\delta}$. As seen from Eq. (24), the physics is
dominated by the dressed spinon self-energy renormalization due to
the dressed holon pair bubble. In this sense, the mobile dressed
holons are the key factor leading to the IC magnetic correlation,
i.e., the mechanism of the IC type of structure in doped cuprates
is most likely related to the dressed holon motion. This is why
the position of the IC peaks can be determined in the present
study within the $t$-$t'$-$J$ model, while the dressed spinon
energy dependence is ascribed purely to the self-energy effects
which arise from the dressed holon-spinon interaction. Since the
values of ${\rm Im}\Sigma_{s}^{(2)}({\bf k},\omega)$ increase with
increasing energies, then all values of $1/{\rm
Im}\Sigma_{s}^{(2)} ({\bf k},\omega)$ are very small at high
energies, which leads to the IC peaks disappearing at high
energies.

\section{Summary and discussions}

In summary, we have developed a partial CSS fermion-spin theory to
study the physical properties of the underdoped cuprates. In this
approach, the physical electron is decoupled completely as the
dressed holon and spinon, with the dressed holon keeps track of
the charge degree of freedom together with the phase part of the
spin degree of freedom, while the dressed spinon keeps track of
the amplitude part of the spin degree of freedom. The electron
local constraint for single occupancy is satisfied in analytical
calculations. The dressed holon is a magnetic dressing, and it
behaves like a spinful fermion, while the dressed spinon is
neither boson nor fermion, but a hard-core boson. Moreover, both
dressed holon and spinon are gauge invariant, and in this sense,
they are real and can be interpreted as physical excitations. In
the common decoupling scheme, we obtain the full dressed holon and
spinon Green's functions by using the equation of motion method.

Within this theoretical framework, we have studied the charge and
spin dynamics of the underdoped cuprates based on the $t$-$t'$-$J$
model. The conductivity spectrum contains a non-Drude low energy
peak and a broad midinfrared band, while the temperature dependent
resistivity is characterized by a crossover from the moderate
temperature metallic-like to the low temperature insulating-like
behavior in the heavily underdoped regime, and a temperature
linear dependence with deviations at low temperatures in the
underdoped regime. The commensurate neutron scattering peak at the
half-filling is split into IC peaks with dopings, where the
incommensurability is doping dependent, and increases with the
hole doping concentration at lower dopings, and saturates at
higher dopings. These results are qualitatively similar to those
seen in experiments. It is essential that these theoretical
results were obtained without any adjustable parameters. These
results also show that the charge dynamics is mainly governed by
the scattering from the dressed holons due to the dressed spinon
fluctuation, while the scattering from the dressed spinons due to
the dressed holon fluctuation dominates the spin dynamics. In this
case, the spin and charge dynamics in the normal-state are almost
independent, and the perturbations that interact primarily with
the charge do not much affect the spin \cite{pwa4}, therefore the
notion of partial CSS naturally accounts for the qualitative
features of the normal-state of the underdoped cuprates.

Based on this partial CSS fermion-spin theory, we have discussed
the mechanism of HTSC in doped cuprates \cite{feng5}. It is shown
that dressed holons interact occurring directly through the
kinetic energy by exchanging the dressed spinon excitations,
leading to a net attractive force between the dressed holons, then
the electron Cooper pairs originating from the dressed holon
pairing state are due to the charge-spin recombination, and their
condensation reveals the superconducting ground-state. The
electron superconducting transition temperature is determined by
the dressed holon pair transition temperature, and is proportional
to the hole doping concentration in the underdoped regime. To our
present understanding, the main reasons why the present theory is
successful in studying the physical properties of doped cuprates
are that (1) the electron local constraint is exactly obeyed
during analytical calculations in contrast with the slave-particle
approach, where the local constraint is explicitly replaced by a
global one \cite{bcmps,feng1}. In this case, the representation
space in the slave-particle approach is much larger than the
representation space for the physical electron. (2) Since the
extra $U(1)$ gauge degree of freedom related with the local
constraint has been incorporated into the dressed holon, this
leads to the fact that the dressed holon and spinon are gauge
invariant in the partial CSS fermion-spin theory. However, the
bare holon and spinon in the slave-particle theory are strongly
coupled by the $U(1)$ gauge field fluctuation, they are not gauge
invariant. (3) The representation of the dressed spinon in terms
of the spin raising and lowering operators is essential in the
present approach \cite{feng1}, because whenever a dressed holon
hops it gives rise immediately to a change of the spin background
as a result of careful treatment of the constraint given in Sec.
II. This is why the dressed holon-spinon interaction (kinetic
part) dominates the essential physics of the underdoped cuprates
\cite{feng2,feng4}.

\acknowledgments
The authors would like to thank Professor Z.Q. Huang, Professor
Z.B. Su, Professor L. Yu, Dr. F. Yuan, and Professor Z.X. Zhao
for the helpful discussions. This work was supported by the
National Natural Science Foundation of China under Grant Nos.
10125415 and 10074007.

\end{document}